\documentstyle[twocolumn,eqsecnum,aps]{revtex}

\def\Tr{\mathop{\rm Tr}}

\begin{document}
\draft
\preprint{}
\title{Numerical Renormalization Approach to Two-Dimensional Quantum
Antiferromagnets with Valence-Bond-Solid Type Ground State
}
\author{ Yasuhiro Hieida, Kouichi Okunishi and Yasuhiro Akutsu}
\address{
Department of Physics, Graduate School of Science, Osaka University,\\
Machikaneyama-cho 1-1, Toyonaka, Osaka 560-0043, Japan.
}
\date{\today}
\maketitle
\begin{abstract}
We study the ground-state properties of the two-dimensional quantum
spin systems having the valence-bond-solid (VBS) type ground
states. The ``product-of-tensors'' form of the ground-state
wavefunction of the system is utilized to associate it with an
equivalent classical lattice statistical model which can be treated by
the transfer-matrix method.  For diagonalization of the transfer
matrix, we employ the product-wavefunction renormalization group
method which is a variant of the density-matrix renormalization group
method.  We obtain the correlation length and the sublattice
magnetization accurately.  For the
 anisotropically ``deformed''
 $S=3/2$ VBS model
on the honeycomb lattice, we find that the correlation length as a
function of the deformation parameter behaves very much alike as that
in the $S=3/2$ VBS chain.
\end{abstract}

\narrowtext

\section{Introduction}
\label{sec1}

There are many applications of the density matrix renormalization
group (DMRG)~\cite{orig-DMRG1,orig-DMRG2} which was originally applied
to the one-dimensional (1D) quantum
systems~\cite{orig-DMRG1,orig-DMRG2}. Due to its remarkable success,
the DMRG has now become one of the standard methods for studying 1D
quantum models, two-dimensional (2D) classical models~\cite{Nishino}
and $(1+1)$-dimensional classical non-equilibrium
models~\cite{Hieida,KP}.

As was pointed out in Refs.~\cite{OR-1,OR-2}, DMRG is a variational
method under the matrix-product-form ansatz (MPFA) for trial
wavefunctions whose usage dates back to the work of Kramers and
Wannier~\cite{KW,Nishino-Okunishi-CTMRG-1,Nishino-Okunishi-CTMRG-2}.
This point of view leads to some non-trivial reformulations of the
DMRG: the direct variational approach,~\cite{OR-1,OR-2,Kennedy,SZ-MP}
the product-wavefunction renormalization group (PWFRG),~\cite{cPWFRG}
the corner-transfer-matrix renormalization group
(CTMRG).~\cite{Nishino-Okunishi-CTMRG-1,Nishino-Okunishi-CTMRG-2}
Further, in this view, the success of the DMRG implies the unexpected
accuracy of the MPFA wavefunctions.

Having seen the success of the DMRG for 1D quantum and 2D classical
systems, it is natural and important to explore its higher-dimensional
generalizations.  For 2D quantum case, the ladder
approach,~\cite{White-2dQ,2dITF,Sierra} can be regarded as such, but
it is essentially the one-dimensional algorithm.  For 3D classical
case, as an extension of the
CTMRG,~\cite{Nishino-Okunishi-CTMRG-1,Nishino-Okunishi-CTMRG-2} the
``corner-tensor'' approach has recently been
proposed,~\cite{Nishino-3dIsing} which is the first among the truly
higher-dimensional algorithms, although it may not be the only
possibility.

In our view, the most promising one in generalizing the DMRG to higher
 dimensions seems to be the one which is based on a generalization of
 the MPFA.  We should remark that, we have already known a one: a
 wavefunction which is a product of local tensors, typical example
 being the valence-bond-solid (VBS) state~\cite{AKLT-1,AKLT-2} for a
 class of 2D quantum antiferromagnets. It has been well known that the
 wavefunction of the 1D VBS state can be expressed as a product of
 $2\times2$ matrices, implying that the MPFA is exact for the system;
 it has also been known that the wavefunction of the 2D (or higher
 dimensional) VBS state can be expressed as a product of ``tensors''
 (generalized objects of matrices).  Let us simply call such form of
 wavefunction a tensors-product-form (TPF) wavefunction.  A natural
 generalization of the MPFA is, then, the tensors-product-form ansatz
 (TPFA).  Accordingly, we can formulate the TPFA-variational method
 which can be thought of a ``higher-dimensional DMRG''.  The aim of
 the present article is to make the first step in this
 TPFA-variational approach to higher-dimensional quantum systems.  We
 focus on the property of the VBS-type state in 2D, and for this
 purpose, we develop a reliable numerical method at the same time.

The models we consider in the present article is the 2D quantum
 antiferromagnets which have the VBS-type ground
 states;~\cite{AKLT-1,AKLT-2,NKZ} the TPFA
 is
 exact for these models.
 Let $|{\rm VBS}\rangle$ be the unnormalized ground-state vector whose
 wavefunction is exactly given by a product of local tensors.  A main
 part of our problem is to evaluate the expectation value of a given
 observable ${\cal O}$
\begin{equation}
\langle {\cal O}\rangle
 \equiv \langle {\rm VBS}|{\cal O}|{\rm VBS}\rangle
        /\langle {\rm VBS}|{\rm VBS}\rangle. 
\label{expectation} 
\end{equation}
For 1D VBS-type models, due to the matrix-product-form structure of
 $|{\rm VBS}\rangle$, the RHS of (\ref{expectation}) can be
 interpreted~\cite{AKLT-1,AKLT-2} as a thermal average in a 1D
 classical statistical-mechanical model; the transfer-matrix method
 allows us to evaluate the RHS of (\ref{expectation}) exactly.  For 2D
 VBS-type models, similar interpretation as a 2D classical
 statistical-mechanical problem is straightforward due to the TPF
 structure of $|{\rm VBS}\rangle$.  Based on this interpretation,
 Niggemann {\em et al.}  made the {\em classical} Monte-Carlo study of
 the anisotropically generalized $S=3/2$ VBS model on the honeycomb
 lattice.  The approach we take in the present work is, in a sense, a
 direct generalization of the 1D case; we treat the associated 2D
 classical statistical-mechanical problem by the transfer-matrix
 method.  What is essential in our approach is that, for
 diagonalization of the transfer matrix, we employ the DMRG (to be
 precise, PWFRG, in the present article) allowing us to make highly
 reliable, close-to-exact evaluation of (\ref{expectation}).  The use
 of the DMRG also has an important implication, in the light of the
 TPFA-variational method: 2D TPFA-variational calculation is reduced
 to 1D TPFA-variational method, namely the DMRG.  Accordingly, we can,
 in principle, formulate the ``nested'' TPFA-variational approach
 where $D$-dimensional TPFA-variational calculation is reduced to
 $(D-1)$-dimensional one, which in turn is reduced to
 $(D-2)$-dimensional one, and so on.

This paper is organized as follows. In Section~\ref{sec2}, we give
explicit form of local tensors for the 2D VBS-type models.  For actual
calculations, we consider the $S=2$ isotropic VBS model on the square
lattice, and the ``deformed'' $S=3/2$ VBS-type model on the honeycomb
lattice proposed by Niggemann {\em et al}. In Section~\ref{sec3}, we
explain our calculation method.  We extend the PWFRG so that it can
handle the asymmetric transfer matrices which we encounter in treating
the 2D VBS-type models. In Section~\ref{sec4}, calculated results are
given.  The last section (Section~\ref{sec5}) is devoted to summary
and conclusion.

\section{Vertex Model Associated with the Valence-Bond-Solid States}
\label{sec2}

\subsection{Two-Dimensional Valence-Bond-Solid Wavefunction and Local Tensors}

Construction of the higher dimensional VBS state has already been made
in the original paper.\cite{AKLT-1} We shall give a brief explanation
of the TPF structure of the general VBS state.

Consider a spin-$S$ operator situated at a site $i$ on a lattice which
may not necessarily be a ``regular'' one.  Let $\{|\sigma\rangle\}$
($\sigma=-S,-S+1,\ldots,S$) be the corresponding $s^{z}$-diagonal
basis set. We assume that the site $i$ is ``$M$-valent'' with $M=2S$.
We then prepare $M$ spin-1/2 operators, and regard the spin-$S$
operator as the one constructed from these component operators.  Let
$|\eta_{\alpha}\rangle$ ($\eta_{\alpha}=\pm1/2$, $\alpha=1,\ldots,M$)
be the $s^{z}$-diagonal base of the $\alpha$-th component spin, and
denote
\begin{equation}
|\{\eta\}\rangle\equiv|\eta_{1},\eta_{2},\ldots,\eta_{M}\rangle \equiv
 |\eta_{1}\rangle\otimes|\eta_{2}\rangle\otimes\cdots\otimes|\eta_{M}\rangle .
\end{equation}
Consider the symmetrized state
\begin{equation}
|(\{\eta\})\rangle
 \equiv
 |(\eta_{1},\eta_{2},\ldots,\eta_{M})\rangle
\equiv
 {\cal N}_{\eta}{\rm Sym_{\eta}} |\eta_{1},\eta_{2},\ldots,\eta_{M}\rangle, 
\end{equation}
where ${\rm Sym}_{\eta}$ stands for the symmetrization with respect to
the indices $\{\eta_{\alpha}\}$, and ${\cal N}_{\eta}$ is a factor
introduced for the proper normalization,
$\langle(\{\eta\})|(\{\eta\})\rangle=1$.  Clearly, we have
\begin{equation}
|\sigma\rangle
 =
 |(\{\eta\})\rangle ,
 \quad\mbox{for $\sigma=\sum_{\alpha}\eta_{\alpha}$}.
\end{equation}
We then define a local tensor
 $A(\sigma|\{\eta\})$($=A(\sigma|\eta_1,\eta_2,\ldots,\eta_{M})$) by
\begin{equation}
A(\sigma|\{\eta\})\equiv \langle \sigma|(\{\eta\})\rangle,
\label{VBSweight}
\end{equation}
which is the building block of the VBS state (see Fig.~\ref{fig:A}).

\fbox{fig-1}

 Note that, by definition, the tensor
$A(\sigma|\eta_1,\eta_2,\ldots,\eta_{M})$ is symmetric with respect to
the indices $\{\eta_{\alpha}\}$.  Further, we adopt the phases of the
vectors $\{|\sigma\rangle\}$ so that the tensor has the property
\begin{equation}
A(\sigma|\{\eta\})=A(-\sigma|\{-\eta\}). \label{negsymm}
\end{equation}

Let $A_{i}(\sigma_{i}|\{\eta^{i}\})$ be the local tensor associated
with the site $i$.  The VBS wavefunction $\Psi_{\rm
VBS}(\sigma_{1},\sigma_{2},\ldots,\sigma_{N})$ is given by
\begin{equation}
\Psi_{\rm VBS}(\sigma_{1},\sigma_{2},\ldots,\sigma_{N})
=\sum_{\eta}
 \prod_{<i,j>}
 S_{ij}(\eta^{i},\eta^{j})
 \prod_{i}A_{i}(\sigma_{i}|\{\eta^{i}\}), \label{VBSwf}
\end{equation}
where we have introduced the sign factor $S_{ij}(\eta^{i},\eta^{j})$
for each connected site pair $<i,j>$, corresponding to formation of
the ``valence bond''.\cite{AKLT-1,AKLT-2} For the valence bond formed
from the $\alpha$-th bond at the site $i$ and $\beta$-th bond at the
site $j$, the sign factor is explicitly given by
\begin{equation}
S_{ij}(\eta^{i},\eta^{j})=\epsilon(\eta^{i}_{\alpha},\eta^{j}_{\beta}),
\end{equation}
where
\begin{equation}
\epsilon(\xi,\xi')=\left\{
\begin{array}{ll}
1 & \mbox{($\xi=-1/2$ and $\xi'=1/2$)}\\
-1 & \mbox{($\xi=1/2$ and $\xi'=-1/2$)}\\
0 & \mbox{(otherwise)}\\
\end{array}
\right.
.
\end{equation}
The sign factor can be absorbed into the definition of modified local
tensor as
\begin{equation}
\bar{A}(\sigma|\eta_{1},\ldots,\bar{\eta}_{\alpha},\ldots,)
=\sum_{\eta_{\alpha}'}
 \epsilon(\bar{\eta}_{\alpha},\eta_{\alpha}')
 A(\sigma|\eta_{1},\ldots,\eta_{\alpha}',\ldots,)
\label{eq:A-bar}
\end{equation}
We can introduce modified tensors where two or more indices (at
$\alpha$-th, $\beta$-th, $\ldots$, positions) are replaced by those
with bars ($\bar{A}(\sigma|
\eta_{1},\ldots,\bar{\eta}_{\alpha},\ldots,\bar{\eta}_{\beta},\ldots)$,
etc.).  In terms of the modified tensors, we can express the VBS
wavefunction as a product of these tensors summed over the
bond-variables $\{\eta\}$ in a form similar to (\ref{VBSwf}), with the
sign factor $S_{i,j}(\eta^{i},\eta^{j})$ simply replaced by the
Kronecker's delta
\begin{equation}
\delta(\eta^{i}_{\alpha},\bar{\eta}^{j}_{\beta}),
\end{equation}
where we have assumed that the $\alpha$-th bond at site $i$ and
$\beta$-th bond at site $j$ are to be connected, and that the tensor
at site $j$ is of the modified type at the $\beta$-th position.

\subsection{Vertex-Model Interpretation}

We can regard the tensor $A(\sigma|\{\eta\})$ as the Boltzmann weight
of a statistical-mechanical model where both the spin variables
$\{\sigma_{i}\}$ (``vertex spins'') and the bond variables
$\{\eta^{i}_{\alpha}\}$ are fluctuating variables, which we may call
spin-vertex model.  Then the unnormalized VBS wavefunction
(\ref{VBSwf}) is the partition function of the spin-vertex model with
fixed vertex-spin configuration
$\{\sigma_{1},\sigma_{2},\ldots,\sigma_{N}\}$.

To obtain the quantum-mechanical expectation (\ref{expectation}), we
should calculate the norm $\langle\mbox{VBS}|\mbox{VBS}\rangle$ which
can also be regarded as a partition function of a lattice model,
namely, the vertex model.  Assuming that the local tensors
$\{A_{i}(\sigma_{i}|\eta^{i})\}$ are real, we can write the vertex
weight for the $M$-valent site as
\begin{equation}
W[(\eta_{1},\xi_{1}),\ldots,(\eta_{M},\xi_{M})]
=\sum_{\sigma} A(\sigma|\{\eta\})A(\sigma|\{\xi\}).
\label{eq:cWeight-1}
\end{equation}
The sign factor in (\ref{VBSwf}) can also be taken into account by
introducing the modified vertex weights, for example,
\begin{eqnarray}
&&\bar{W}[\ldots,(\bar{\eta}_{\alpha},\bar{\xi}_{\alpha}),\ldots]\nonumber\\
&&=\sum_{\eta_{\alpha}',\xi_{\alpha}'}
   \epsilon(\bar{\eta}_{\alpha},\eta_{\alpha}')
   \epsilon(\bar{\xi}_{\alpha},\xi_{\alpha}')
   W[\ldots,(\eta_{\alpha}',\xi_{\alpha}'),\ldots].
\label{eq:cWeight-2}
\end{eqnarray}
In this vertex model, the double index $(\eta_{\alpha},\xi_{\alpha})$
($\alpha=1,2,\ldots$) plays the role of the bond variable in the
ordinary vertex model.  Since each component index takes two values
$\pm 1/2$, the model is a $4$-state vertex model. The
quantum-mechanical expectation (\ref{expectation}) can then be
regarded as a statistical-mechanical average in the vertex model.

\subsection{Square and Honeycomb Lattices}

We treat the VBS-state-associated vertex model by the transfer matrix
method.  For ease of construction of the row-to-row transfer matrix,
we restrict the analysis to the two cases in the present paper: $S=2$
VBS model on the square lattice and the $S=3/2$ VBS model on the
honeycomb lattice.  In the former, the row-to-row transfer matrix can
be constructed in the conventional way.\cite{Baxter} In the latter, we
can map the model on the honeycomb lattice to the one on the square
lattice\cite{NKZ} as shown in Fig.~\ref{fig:map}.  Hence either case
can be treated as a vertex models on the square lattice.

\fbox{fig-2}

From the general formula (\ref{VBSweight}), we can explicitly write
down non-zero components of the local tensor.  For $S=3/2$
honeycomb-lattice VBS model, we have
\begin{eqnarray}
A(3/2|1/2,1/2,1/2)  &=&1, \nonumber\\
A(1/2|-1/2,1/2,1/2) &=&\frac{1}{\sqrt{3}}, \label{s=3/2weight}
\end{eqnarray}
and for $S=2$ square-lattice VBS model, we have
\begin{eqnarray}
A(2|1/2,1/2,1/2,1/2)  &=&1, \nonumber\\
A(1|-1/2,1/2,1/2,1/2) &=&\frac{1}{2}, \nonumber\\
A(0|-1/2,-1/2,1/2,1/2)&=&\frac{1}{\sqrt{6}}.  \label{s=2weight}
\end{eqnarray}
Other non-zero components are obtained by using the permutational
symmetry of $A(\sigma|\{\eta\})$ in $\{\eta\}$ and the property
(\ref{negsymm}).  The VBS states constructed from these tensors are
exact ground states of the Hamiltonian where the local
nearest-neighbor (nn) -type pair Hamiltonian $h_{ij}$ is the
projection operator into the local spin-$2S$ space, $P^{2S}_{ij}$ ,
($S=3/2$ for honeycomb lattice, and $S=2$ for square
lattice):\cite{AKLT-1,AKLT-2}
\begin{equation}
P^{2S}_{ij}|\mbox{VBS}\rangle =0 , \quad\mbox{for all nn site pair $ij$}.
\end{equation}

\subsection{``Deformed'' VBS-type Model on the Honeycomb Lattice}

In the original VBS model, the local Hamiltonian $P^{2S}_{ij}$ is a
function of the inner product $\vec{s}_{i}\cdot\vec{s}_{j}$, hence, is
isotropic.\cite{AKLT-1,AKLT-2} Anisotropic generalization of the VBS
models and their associate ground-state vector has been
known.\cite{S1deformed,NZ} Similar generalization has also been known
for the honeycomb VBS model.\cite{NKZ} The generalized honeycomb VBS
model contains a ``deformation'' parameter which corresponds to the
$xxz$-type anisotropy of the Hamiltonian.  The associated wavefunction
is also of the TPFA form with a
deformed local tensor
$A(\sigma|\{\eta\})_{a}$ where $a$ is the deformation
parameter. Explicit form of $A(\sigma|\{\eta\})_{a}$ is simply given
by that of the isotropic case (\ref{s=3/2weight}) with substitution
\begin{equation}
\frac{1}{\sqrt{3}} \rightarrow 1/a.
\end{equation}
By $\Psi_{{\rm VBS}(a)}(\sigma_{1},\sigma_{2},\ldots,\sigma_{N})$, we
denote the TPFA wavefunction made from
$\{A_{i}(\sigma_{i}|\{\eta^{i}\})_{a}\}$, and by
$|\mbox{VBS}(a)\rangle$, corresponding state vector.  Since
$|\mbox{VBS}(-a)\rangle$ is essentially equivalent to
$|\mbox{VBS}(a)\rangle$ (via a $S^{z}$-diagonal unitary
transformation), we treat $a>0$ in the present article.
(We have confirmed that calculated results for $-a$ are identical to
those for $a$.)
 As has been
shown in Ref.\cite{NKZ} there actually exists Hamiltonian with nn-type
interactions whose ground state is precisely $|\mbox{VBS}(a)\rangle$
($\times$ normalization factor).

\narrowtext

\section{Calculation Method}
\label{sec3}

In the previous section, we explained how to translate physical
quantities from the quantum-mechanical frame (Eq.~(\ref{expectation})
) into the classical statistical-mechanical one
(Eq.~(\ref{eq:cWeight-1}) and/or Eq.~(\ref{eq:cWeight-2})).

In this section, to deal with the classical statistical-mechanical
 frame, we introduce a transfer matrix.  So now we focus on a way of
 renormalization of an {\it asymmetric} transfer matrix by using the
 ``classical PWFRG''~\cite{cPWFRG}.  Our transfer matrix is composed
 of 4-state vertex weights.  We denote this 4-state vertex weight by
 $E^p_q (a,b)$, which is depicted by Fig.~\ref{fig:E}.  We regard this
 $E^p_q (a,b)$ as a matrix element of $E^p_q$.

\fbox{fig-3}

\subsection{Magnetization}
Let us explain how to obtain the expectation value of any one-point
operator $A_i$ at site $i$ (by the PWFRG):
\begin{equation}
\langle A_i \rangle 
  = \frac{\langle \psi | A_i |\psi \rangle}
         {\langle \psi |\psi \rangle},
\label{eq:mag}
\end{equation}
where $\psi$ represents the ground state of the 
 VBS model (for instance, $|\psi \rangle=|\mbox{VBS}(a)\rangle$).
We represent the numerator of Eq.~(\ref{eq:mag}) by transfer matrices
$T$ and $T_{\rm A}$, whose definitions will be shown later
 (Eq.~(\ref{eq:defTA}) and Eq.~(\ref{eq:Tid})):
\begin{equation}
\langle \psi | A_i |\psi \rangle
= \Tr{\left [T_A (T)^{N-1}\right ]},
\label{eq:nume-0}
\end{equation}
where $N$ is a linear size.  $T$ and $T_{\rm A}$ in this equation are
defined in the following (for short , we represent $\{\ldots
p_{-1},p_{0},p_{1},\ldots \}$ by $\{p\}$).
\begin{equation}
\left [T_{\rm A}\right ]^{\{p\}}_{\{q\}}
\equiv
 \left [ 
        \prod_{j=-(M-1)}^{-1} E^{p_j}_{q_j}
 \right ]
 \left [
  E_{\rm A}
 \right ]^{p_0}_{q_0}
 \left [
  \prod_{j=1}^{M} E^{p_j}_{q_j}
 \right ]
\label{eq:defTA}
\end{equation}
where $2M$ is another linear
 size (along this direction, we do
 not impose a periodic boundary condition).  And $E_{\rm A}$ is
 composed
 of a quantum-mechanical operator $A$ which is sandwiched
 between classical weights 
(Eq.~(\ref{VBSweight}) and/or Eq.~(\ref{eq:A-bar})).
 Note that $E\equiv E_{\rm id}$ ({\rm id}: a local identity operator).
Using these notations,
\begin{equation}
T_{\{p\},\{q\}}
\equiv \left [T\right ]^{\{p\}}_{\{q\}}
\equiv \left [T_{\rm id}\right ]^{\{p\}}_{\{q\}}
\equiv \prod_{j=-(M-1)}^{M} E^{p_j}_{q_j} .
\label{eq:Tid}
\end{equation}

Furthermore $T$ is decomposed into
\begin{equation}
T = R D L,
\label{eq:diag}
\end{equation}
 where $R$ is composed of column right eigenvectors, and $D$ is
diagonal matrix whose diagonal elements $\{\lambda_{j}\}$ (these
$\lambda_{1},\lambda_{2},\ldots $ are in descending order of the
absolute value) is the set of eigenvalues of the transfer matrix $T$
and $L$ is consist of row left eigenvectors.  Here we should keep in
mind that $T$ is, generally, an asymmetric matrix, upon whose left and
right eigenvectors $L$ and $R$ we impose the normalization condition
\begin{equation}
LR=1.
\label{eq:LR}
\end{equation}

Using Eqs.~(\ref{eq:nume-0}) and (\ref{eq:diag}) and (\ref{eq:LR}),
\begin{eqnarray}
& &\langle \psi | A_i |\psi \rangle \nonumber \\
&=&\sum_{j}(L T_A R)_{j,j}(\lambda_{j})^{N-1}\nonumber\\
&\cong&(L T_{\rm A} R)_{1,1}(\lambda_{1})^{N-1} \nonumber \\
&=&\langle l |T_A|r\rangle (\lambda_{1})^{N-1},
\label{eq:nume}
\end{eqnarray}
 where the subscript ``1'' represents the eigenstate with the largest
eigenvalue of $T$ in absolute value and we denote its left eigenvector
and right eigenvector by $\langle l|$ and $|r\rangle$ respectively.

The denominator of Eq.~(\ref{eq:mag}) (the ``partition function'')
is obtained by
\begin{equation}
\langle \psi |\psi \rangle 
=\Tr{\left [ T^N \right ]}
\cong\lambda_1^N.
\label{eq:deno}
\end{equation}

From Eqs.~(\ref{eq:mag}), (\ref{eq:nume}) and (\ref{eq:deno}),
we get one-point function
\begin{equation}
\langle A_i \rangle 
=\frac{\langle l|T_{\rm A}|r\rangle }{\lambda_{1}}.
\label{eq:mag-2}
\end{equation}

To get $\langle l|$, $T_{\rm A}$, $|r\rangle $ and $\lambda_1$ in this
equation, we use the PWFRG method (infinite method).  As we stated
earlier, this transfer matrix is {\it asymmetric}.  So we decide how
to renormalize the asymmetric transfer matrix using the PWFRG.  This
is because there are several ways to treat an asymmetric transfer
matrix.  In this paper, we treat it in the following way which is
different from our previous treatment.~\cite{Hieida} In our previous
treatment, we adopted asymmetric density matrices in the DMRG.  Unlike
this, in this paper, our treatment is equivalent to treating {\it
symmetric} density matrices in the PWFRG (namely, we use the singular
value decomposition (SVD) for left and right eigenvectors of $T$).
It is easier to perform the renormalization with the SVD than
renormalization with diagonalization of asymmetric matrices, because
the former is free from difficulties of manipulating complex numbers.
In compensation for this difficulty, we only have to introduce
identity operators as will be shown later (see Eq.~(\ref{eq:PowerMethod})). 
These identity operators is very important for the correct
renormalization of an asymmetric transfer matrix.

First of all, we present the outline of our prescription in this paper
before explaining details.

Our prescription in this paper is:
\vspace*{2em}
\begin{description}
\item[P-1 ] 
Multiply left eigenvector and right eigenvector obtained in the
 previous iteration by transfer matrix (and by an identity operator)
 several times to get the improved left eigenvector and right
 eigenvector.

\item[P-2 ]
 Perform SVD for the improved left and right eigenvectors,
 respectively to get
 ``projectors''.

\item[P-3 ]
 Select important states and renormalize transfer matrices,
 identity operators and one-point operators.

\item[P-4 ]
 Get improved ``projectors'' using the ``recursion
 relation''~\cite{cPWFRG} and, from these ``projectors'', construct new
 left and right eigenvectors.

\item[P-5 ]
Evaluate 
one-point functions such as
 $\langle A_i \rangle $ by
using Eq.~(\ref{eq:mag-2}).
 Check the convergence of both left and right
 wavefunctions and the value of the one-point function.  If convergence is
 true then we calculate 2-point function or evaluate $\lambda_2$ to
calculate the correlation length (see section~\ref{sec:corr}).

\item[P-6 ]
 Return step {\bf P-1}.
\end{description}

\vspace*{2em}

The followings are the details in the above prescription:

\noindent
(Hereafter, we use Greek indices to represent the block state.)

\vspace*{1em}

First, a new notation is introduced. We renormalize bare $T$ into 
renormalized transfer matrix
 $\widetilde{T}$
\begin{equation}
\widetilde{T}^{\alpha,i,j,\beta}_{\alpha',i',j',\beta'}
\equiv \sum_{k=1}^4
 \left [{\tt T}_{\rm L}\right ]^{\alpha,i}_{\alpha',i'}(k)
 \left [{\tt T}_{\rm R}\right ]^{j,\beta}_{j',\beta'}(k),
\label{eq:renT}
\end{equation}
where, in this equation, ${\tt T}_{\rm L}$ and ${\tt T}_{\rm R}$ are
renormalized as Eq.~(\ref{eq:TLnew}).  Eq.~(\ref{eq:renT}) is
expressed according to Nishino's diagram~\cite{Nishino} in
Fig.~\ref{fig:renT}.

\fbox{fig-4}

Second, we give caveats in our prescription below:

As for 
{\bf P-1}:\\
To concretely explain what is mentioned in {\bf P-1}, we take the case
of improving right eigenvector
 $|r \rangle_{\rm imp}$.
  It is
important that whenever we multiply by a transfer matrix, we
simultaneously {\it must} multiply by
left and right renormalized identity operators
${\tt id}_{\rm L}$ and ${\tt id}_{\rm R}$.
  Assuming that $k$ is some integer, 
\begin{equation}
|r \rangle_{\rm imp}
=
\left (
\left ({\tt id}_{\rm L}\right )^T
\left ({\tt id}_{\rm R}\right )^T
\widetilde{T}
\right )^k|r\rangle,
\label{eq:PowerMethod}
\end{equation}
where $\left ({\tt id}_{\rm L}\right )^T$ is the transpose of
 ${\tt id}_{\rm L}$ etc.
More concretely, rewriting this equation in its component
 (we treat the case $k=1$ for notation simplicity),
\begin{eqnarray}
& &\langle \mu, i, j, \nu|r\rangle_{\rm imp}=\nonumber \\
&&\sum_{\alpha,\beta,\alpha',i',j',\beta' }
\left [{\tt id}_{\rm L}\right ]^\alpha_\mu \;
 \left [{\tt id}_{\rm R}\right ]^\beta_\nu
\widetilde{T}^{\alpha,i,j,\beta}_{\alpha',i',j',\beta'}
\langle \alpha',i',j',\beta' |r\rangle,
\end{eqnarray}
where 
 $\left [{\tt id}_{\rm L}\right ]^\alpha_\mu$ and
 $\left [{\tt id}_{\rm R}\right ]^\beta_\nu$ are
 left and right renormalized identity operators
 (see Eq.~(\ref{eq:ridL})).

As for
 {\bf P-2}:\\
SVD is performed as:
\begin{equation}
  \langle l |\alpha,i,j,\beta \rangle
   = \sum_{\mu}
      \left [{\tt U}_{\rm L}\right ]^{\alpha,i}_\mu \;
      \left [\sigma_{\rm L} \right ]_\mu \;
      \left [{\tt V}_{\rm L}\right ]^{j,\beta}_\mu,
\end{equation}
  and  
\begin{equation}
  \langle \alpha,i,j,\beta|r\rangle
   = \sum_\mu
     \left [{\tt U}_{\rm R}\right ]^{\alpha,i}_{\mu}\;
     \left [\sigma_{\rm R}\right ]_\mu \;
     \left [{\tt V}_{\rm R}\right ]^{j,\beta}_\mu,
\end{equation}
  where $\{\left [\sigma_{\rm L}\right ]_\mu \}_\mu$ and
        $\{\left [\sigma_{\rm R}\right ]_\mu \}_\mu$
 are
 sets of singular values. And ${\tt U}_{\rm L}$, ${\tt V}_{\rm L}$,
${\tt U}_{\rm R}$ and ${\tt V}_{\rm R}$ are orthogonal matrices.

As for
 {\bf P-3}:\\
We present examples.
New renormalized transfer matrix:
${\tt T}_{\rm L}^{({\rm new})}$
 is given by
\begin{equation}
\left [{\tt T}_{\rm L}^{(\rm new)}\right ]^{\mu,i}_{\nu,j}(l)
=\sum_{\alpha,\beta,k}
\left [{\tt U}_{\rm L}\right ]^\alpha_\mu \;
\left [{\tt T}_{\rm L}\right ]^\alpha_\beta (k)
 E^i_j (k,l) \;
\left [{\tt U}_{\rm R}\right ]^\beta_\nu .
\label{eq:TLnew}
\end{equation}

And the left identity operator: ${\tt id}_{\rm L}$ is
 renormalized as
\begin{equation}
\left [{\tt id}_{\rm L}^{({\rm new})}\right ]^{\alpha}_{\beta}
=\sum_{\alpha^\prime,\beta^\prime,i}
\left [{\tt U}_{\rm L}\right ]^{\alpha',i}_{\alpha} \;
\left [{\tt id}_{\rm L}\right ]^{\alpha'}_{\beta'} \;
\left [{\tt U}_{\rm R}\right ]^{\beta',i}_\beta .
\label{eq:ridL}
\end{equation}
At the very first iteration 
(when we treat a 4-site chain of vertices $T=EEEE$),
 we set
\begin{equation}
\left [{\tt id}_{\rm L}\right ]^\alpha_\beta=\delta_{\alpha,\beta} .
\end{equation}
This is an initial condition for
 the
 recursion relation
Eq.~(\ref{eq:ridL}).

Next, we prepare an object ${\tt TA}_{\rm L}$.  This is for
Eq.~(\ref{eq:OPE}).  In the same way as Eq.~(\ref{eq:renT}), we
renormalize bare $T_{\rm A}$ into
 $\widetilde{T_{\rm A}}$
 as follows.
\begin{equation}
\widetilde{T_{\rm A}}^{\alpha,i,j,\beta}_{\alpha',i',j',\beta'}
\equiv \sum_{k=1}^4
 \left [{\tt TA}_{\rm L}\right ]^{\alpha,i}_{\alpha',i'}(k)
 \left [{\tt T}_{\rm R}\right ]^{j,\beta}_{j',\beta'}(k),
\end{equation}
where ${\tt TA}_{\rm L}$ is made of
\begin{equation}
\left [{\tt TA}_{\rm L}\right ]^{\mu,i}_{\nu,j}(l)
=\sum_{\alpha,\beta,k}
\left [{\tt U}_{\rm L}\right ]^\alpha_\mu \;
\left [{\tt T}_{\rm L}\right ]^\alpha_\beta (k)
\left [E_{\rm A}\right ]^i_j (k,l) \;
\left [{\tt U}_{\rm R}\right ]^\beta_\nu .
\end{equation}

As for
 {\bf P-4}:\\
For example,
 ${\tt U}_{\rm R}$ 's recursion relation is:
\begin{equation}
\left [{\tt U}_{\rm R}^{({\rm new})}\right ]^{\alpha,j}_\mu\nonumber \\
=\sum_{\nu,\beta}
\left [{\tt U}_{\rm R}\right ]^\nu_\alpha
\left [{\tt U}_{\rm R}^{({\rm old})}\right ]^\nu_\beta \;
\left [{\tt U}_{\rm R}\right ]^{\beta,j}_\mu .
\end{equation}

As for
 {\bf P-5}:\\
We show how to obtain the expectation value of any one-point operator
$A_i$ 
 in the next way.
\begin{equation}
\langle A_i \rangle
 =
  \frac{\displaystyle
        \sum_{\alpha,i,j,\beta, \alpha',i',j',\beta'}
        \langle l |\alpha,i,j,\beta \rangle 
        \widetilde{T_{\rm A}}^{\alpha,i,j,\beta}_{\alpha',i',j',\beta'}
        \langle \alpha',i',j',\beta'|r\rangle
       }{
        \displaystyle
        \sum_{\alpha,i,j,\beta, \alpha',i',j',\beta'}
        \langle l |\alpha,i,j,\beta \rangle 
        \widetilde{T}^{\alpha,i,j,\beta}_{\alpha',i',j',\beta'}
        \langle \alpha',i',j',\beta'|r\rangle
        }
\label{eq:OPE}
\end{equation}

\subsection{Correlation Length}
\label{sec:corr}
In what follows, we explain two methods of calculating correlation
lengths. For concreteness, we explain the case of the spin-3/2
honeycomb lattice model. It is also easy to deal with the spin-2
square lattice model.

\subsubsection{Calculus by observation of dumping correlators}
\label{sec:dump-corr}

We set $j-i \equiv r (>0)$ and assume $N \gg r \gg 1$ in the sequel.
And we locate each operator $A$ in the position ``b'' in
Fig.~\ref{fig:map}.

So, we have like Eq.(\ref{eq:nume})
\begin{eqnarray}
& & \langle A_i A_j \rangle \nonumber \\
&=& \frac{\langle \psi |A_i A_j|\psi \rangle}
  {\langle \psi |\psi \rangle} \nonumber \\
&\cong& \sum_{\zeta,\tau}
     (L T_A)_{1,\zeta} [T^{r-1}]_{\zeta,\tau} (T_A R)_{\tau,1}
     (\lambda_1)^{-(r+1)}  \nonumber \\
&=& \langle l|T_A [T^{r-1}] T_A|r\rangle (\lambda_1)^{-(r+1)}.
\label{eq:corr-1}
\end{eqnarray}

From this equation, it follows that we measure correlation length in
the direction depicted in Fig.~\ref{fig:CorrDirec-3}.

\fbox{fig-5}

In off-critical region, we expect 
\begin{equation}
\lim_{j-i\to \infty}
   \langle A_i A_j \rangle
  -\langle A_i \rangle \langle A_j \rangle
  \cong C\exp{(-r/\xi)}/r^{\omega}.
\label{eq:corr-2}
\end{equation}

We bear in mind that in Eq.~(\ref{eq:corr-1}), $\zeta$ and $\tau$
includes $\zeta = 1$ and $\tau = 1$ (let us recall that the subscript
``1'' represents the eigenstate with the largest eigenvalue of $T$ in
absolute value).  So, in the region $\langle A_i \rangle = 0$, we
easily evaluate correlation length from Eq.~(\ref{eq:corr-1}) and
Eq.~(\ref{eq:corr-2}) as
\begin{equation}
\xi^{-1}=\lim_{j-i\to \infty}
\log{\left [
\frac{\langle A_i A_j \rangle}{\langle A_i A_{j+1} \rangle}\right ]}
\end{equation}

\subsubsection{Evaluation by using the power method}

Another method by which
 correlation lengths
 are evaluated is the
following.  By making use of Eq.~(\ref{eq:diag}),
we can further transform Eq.(\ref{eq:corr-1}) into
\begin{equation}
\langle A_i A_j \rangle
\cong \langle A_i \rangle \langle A_j \rangle
  +\tilde{C}_{\rm A}\exp [-r/\xi_{\rm pm}],
\label{eq:corr-p-1}
\end{equation}
where a constant $\tilde{C}_{\rm A}$ is
\begin{equation}
\tilde{C}_{\rm A}
  =\frac{\langle l  |T_{\rm A}|r_2\rangle}{\lambda_2}
   \frac{\langle l_2|T_{\rm A}|r  \rangle}{\lambda_2}
   \frac{\lambda_2}{\lambda_1}
\label{eq:corr-p-2}
\end{equation}
and the definition of $\xi_{\rm pm}$ in this equation is
\begin{equation}
\xi_{\rm pm}^{-1}=
  \log \left | \frac{\lambda_1}{\lambda_2}\right |,
\label{eq:corr-p-3}
\end{equation}
and $\langle l_2|$ and $|r_2\rangle$ are
 the left and right eigenvectors
 of the eigenvalue $\lambda_2$, respectively.

From Eqs.(\ref{eq:corr-p-1}), (\ref{eq:corr-p-2}) and
(\ref{eq:corr-p-3}), we evaluate $\lambda_2$ to obtain the correlation
length $\xi_{\rm pm}^{-1}$.  We use the power method to evaluate
$\lambda_2$ that is, we perform the power method in the space which is
orthogonal to the ground state vectors $\langle l|$ and $|r\rangle$,
as follows.

First, we prepare the initial vectors $\langle \tilde{l}^{(1)}|$ and
$|\tilde{r}^{(1)} \rangle $.  And then we get the improved left
eigenvector $\langle \tilde{l}^{(n)}|$ and the improved right
eigenvector $|\tilde{r}^{(n)}\rangle$.  which are orthogonal to
$\langle l|$ and $|r\rangle$.  Namely,
\begin{equation}
\langle \tilde{l}^{(n)}|
  \equiv
    \langle \tilde{l}^{(n-1)}|T
  - \frac{
      \langle \tilde{l}^{(n-1)}|T|r\rangle}
        {{\langle l|r\rangle}}
    \langle l|,
\end{equation}
and
\begin{equation}
|\tilde{r}^{(n)} \rangle 
  \equiv
     T|\tilde{r}^{(n-1)}\rangle 
  - | r \rangle
    \frac{
      \langle l |T|\tilde{r}^{(n-1)}\rangle}
        {{\langle l|r\rangle}}.
\end{equation}
We should note that our $\langle \tilde{l}^{(n)}|$ and
$|\tilde{r}^{(n)} \rangle $ do not satisfy Eq.(\ref{eq:LR}).  So,
slightly modifying Eq.~(\ref{eq:deno}), we calculate 2nd largest
eigenvalue $\lambda_2$ of $T$ in the following way.
\begin{equation}
\lambda_2=\lim_{n \to \infty} \lambda_2^{(n)},
\end{equation}
where $\lambda_2^{(n)}$ is 
\begin{equation}
\lambda_2^{(n)}=
\frac{
  \langle \tilde{l}^{(n)}|T|\tilde{r}^{(n)}\rangle
     }
     {
      \langle \tilde{l}^{(n)}|\tilde{r}^{(n)}\rangle
     }.
\end{equation}
It should be noted that we save memory resource in a computer by using
the power method (That is we need only $O(m^2)$ memory, not $O(m^4)$,
where $m$ is a number of retained base in the PWFRG).  So, if
necessary, we can perform the above calculation with a larger number
$m$.

\section{Results}
\label{sec4}
\subsection{Isotropic VBS models on honeycomb and square lattices}

As has been mentioned in 
 Sec.~\ref{sec2},
 both the $S=3/2$ VBS model on the honeycomb and the $S=2$ VBS model
on the square lattice can be mapped to classical vertex models on the
square lattice.  We made PWFRG calculation for the sublattice
magnetization and the correlation length in a way as described in
Sec.~\ref{sec3}.

We have confirmed that sublattice magnetization is zero (N\'{e}el order
is absent) in both models; both models are in the disordered phase, in
agreement with previous studies.\cite{AKLT-1,AKLT-2,KLT} As for the
correlation length $\xi$, we obtain
\begin{eqnarray}
\xi^{-1}_{\rm honeycomb}&\approx&1.67 \nonumber\\
\xi^{-1}_{\rm square}&\approx &0.52 \label{xi_isotropic}
\end{eqnarray}
We should note that, in the honeycomb case, the length unit for the
above value of $\xi_{\rm honeycomb}^{-1}$ is 
the lattice spacing of
the mapped square lattice.

\subsection{$S=3/2$ deformed VBS model on the honeycomb lattice}

Unlike the isotropic case, the anisotropic ($xxz$-like) generalization
of honeycomb VBS model can be either in the disordered phase or the
ordered (N\'{e}el) phase, depending on the ``deformation parameter''
$a$.\cite{NKZ}

 The PWFRG allows us to study the behavior of the system in much more
detail, which we present in this subsection. In the actual
calculation, we have observed that the PWFRG calculations are rapidly
convergent in increasing $m$ (number of retained bases), except for
the cases very near the transition point. Consequently, for most
values of $a$, very small $m$, say $m=12$, is sufficient for the
accuracy required in the present study.

\subsubsection{Staggered magnetization}

 In Fig.~\ref{fig:mz}, we show the PWFRG result of the sublattice
magnetization (staggered magnetization) $M_{\rm st}$ as a function of
the parameter $a$.

\fbox{fig-6}

The N\'{e}el order which exists in the large-$a$ region disappear at
$a=a_c\cong 2.54$, in agreement with the Monte-Carlo
result.~\cite{NKZ} It may well be expected that the Ising-like
anisotropy make the system be in the Ising universality class as
regards this anisotropy-induced phase transition.  To check this, we
plot $(M_{\rm st})^8$ versus $a$ as shown in Fig.\ref{fig:mst8}.  The
clear linear behavior near $a_{c}$ implies
\begin{eqnarray}
M_{\rm st} &\sim& (a-a_{c})^{\beta}, \nonumber\\
      \mbox{with}\quad    \beta &=&1/8.
\label{stagmag} 
\end{eqnarray}

\fbox{fig-7}

\subsubsection{Correlation length}
In the transfer-matrix method, the correlation length is usually
calculated from the ratio between the largest and the next-to-largest
eigenvalues of the transfer matrix. This definition of the correlation
length is, in our notation, $\xi_{\rm pm}$ (see
Eq.~(\ref{eq:corr-p-3})), whose PWFRG result is shown in
Fig.\ref{fig:xipower}.

\fbox{fig-8}

  The linear behavior of $\xi^{-1}_{\rm pm}$ near $a_{c}$ is
consistent with the expectation that the system is in the Ising-model
universality-class, namely,
\begin{eqnarray}
\xi^{-1}_{\rm pm}&\sim& |a-a_{c}|^{\nu}\nonumber\\
\mbox{with}\quad \nu &=& 1. \label{xipower}
\end{eqnarray}
In addition to the critical behavior at $a_{c}$, a notable feature of
$\xi_{\rm pm}$ is that it exhibits {\em cusp} singularity at the
isotropic point $a=a_{\rm iso}=\sqrt{3}$.  To clarify the origin of
this cusp, we calculated longitudinal correlation length $\xi_{\rm
l}(a)$ and the transverse correlation length $\xi_{\rm t}(a)$ from the
correlation functions
 $\langle S^z_i S^z_j\rangle$ and $\langle S^x_i S^x_j\rangle$,
 as shown in Fig.\ref{fig:xiinvzx}

\fbox{fig-9}

Totally different $a$-dependence between $\xi_{\rm l}^{-1}(a)$ and
$\xi_{\rm t}^{-1}(a)$ is seen; the isotropic point $a=\sqrt{3}$ is the
crossing point of the two curves.  Since $\xi_{\rm pm}$ is determined
by the correlator with the largest correlation length (or,
equivalently, smallest inverse correlation length), we have
\begin{equation}
\xi_{\rm pm}^{-1}=\left\{
\begin{array}{ll}
\xi_{\rm l}^{-1}(a) & (a\geq\sqrt{3}),\\
\xi_{\rm t}^{-1}(a) & (a<\sqrt{3}),\\
\end{array}
\right.
\end{equation}
which explains the cusp at $a=a_{\rm iso}=\sqrt{3}$ in
Fig.\ref{fig:xipower}.

It is interesting to compare the correlation lengths in the 2D
deformed VBS model with those in the 1D counterpart (deformed $S=3/2$
chain
)~\cite{NZ}, which we denote $\xi_{\rm l}^{(1D)}(a)$ and
$\xi_{\rm t}^{(1D)}(a)$.  Explicitly, we have~{\cite{NZ}}
\begin{eqnarray}
\left (\xi_{\rm l}^{(1D)}\right )^{-1}
  &=&\log \left (\frac{\sqrt{(a^2+1)^2+8}}{|a^2-1|}\right),
\label{eq:1d-xiinv-l}
\\
\left (\xi_{\rm t}^{(1D)}\right )^{-1}
  &=&\log \left (\frac{\sqrt{(a^2+1)^2+8}}{2}\right ),
\label{eq:1d-xiinv-t}
\end{eqnarray}
which also show the crossover behavior at the isotropic point
$a=\sqrt{3}$.  It should be noted that magnitudes of these 1D
correlation lengths are almost the same as those in the 2D case in a
wide range of parameter values.  Major differences are the absence of
the critical point $a_{c}$ and the non-vanishing behavior of $\left
(\xi_{\rm t}^{(1D)}\right )^{-1}$ in the $a\rightarrow0$ limit (see
Fig.~\ref{fig:xiinv}).

\fbox{fig-10}

\begin{figure}
\caption{
Inverse correlation lengths of the free-fermion model (solid curves)
and honeycomb VBS model 
($\circ$:$\xi_{\rm l}^{-1}(a)$, $+$:$\xi_{\rm pm}^{-1}(a)$).
}
\label{fig:ff}
\end{figure}
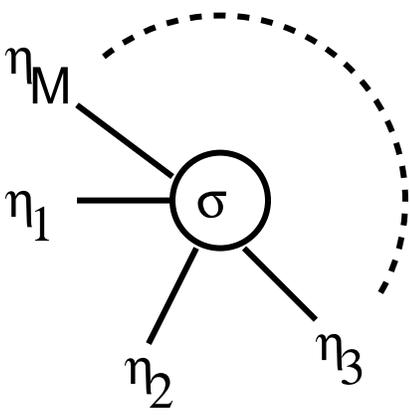
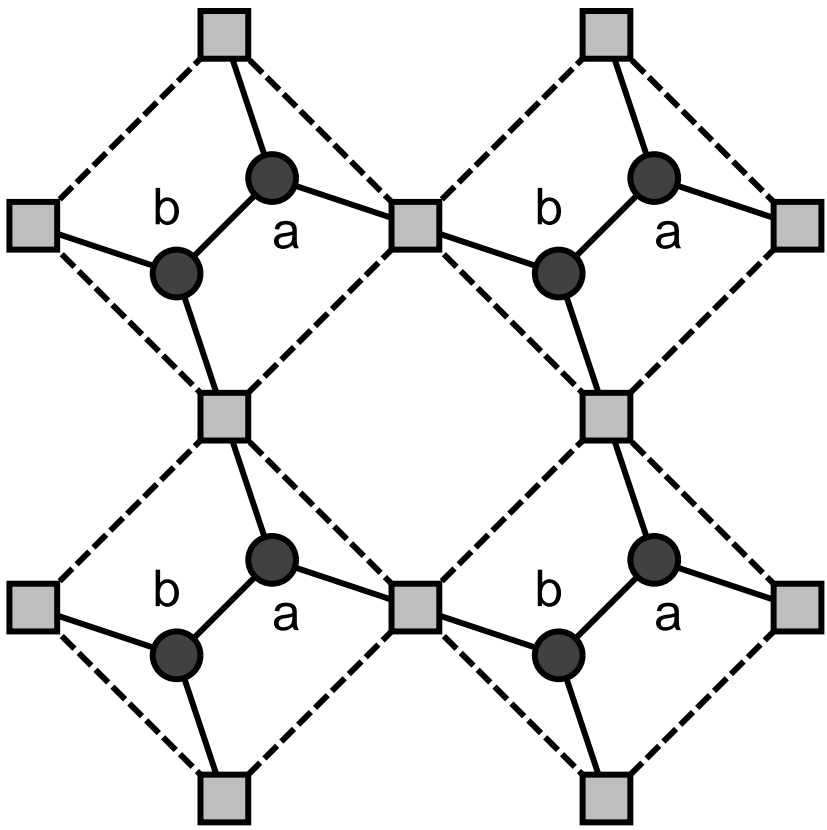
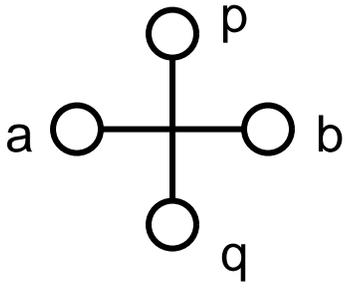
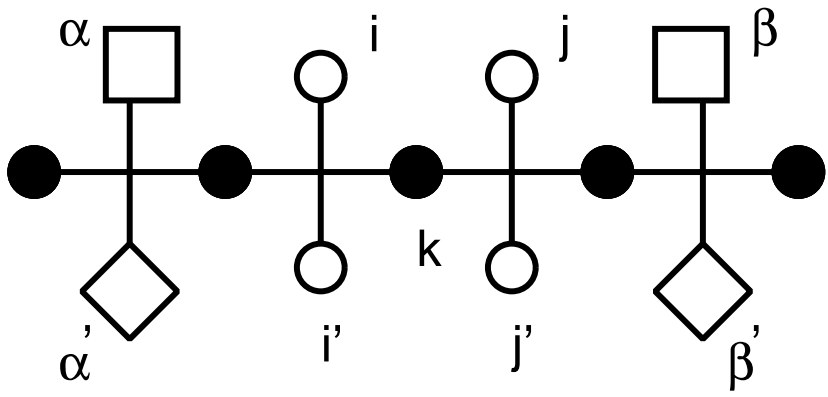
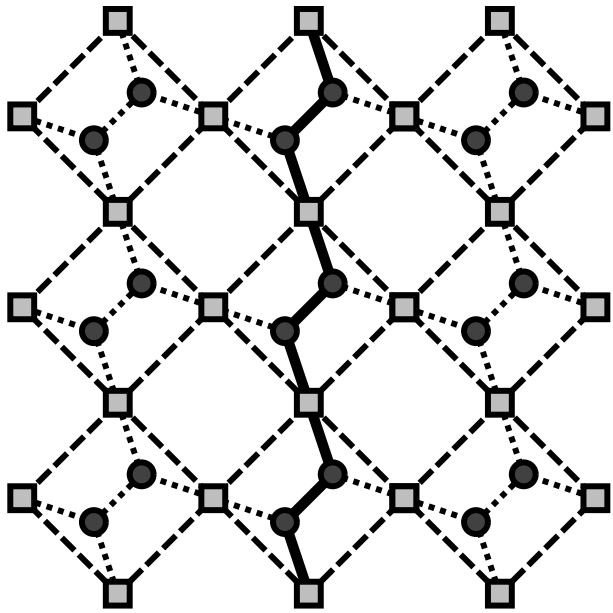
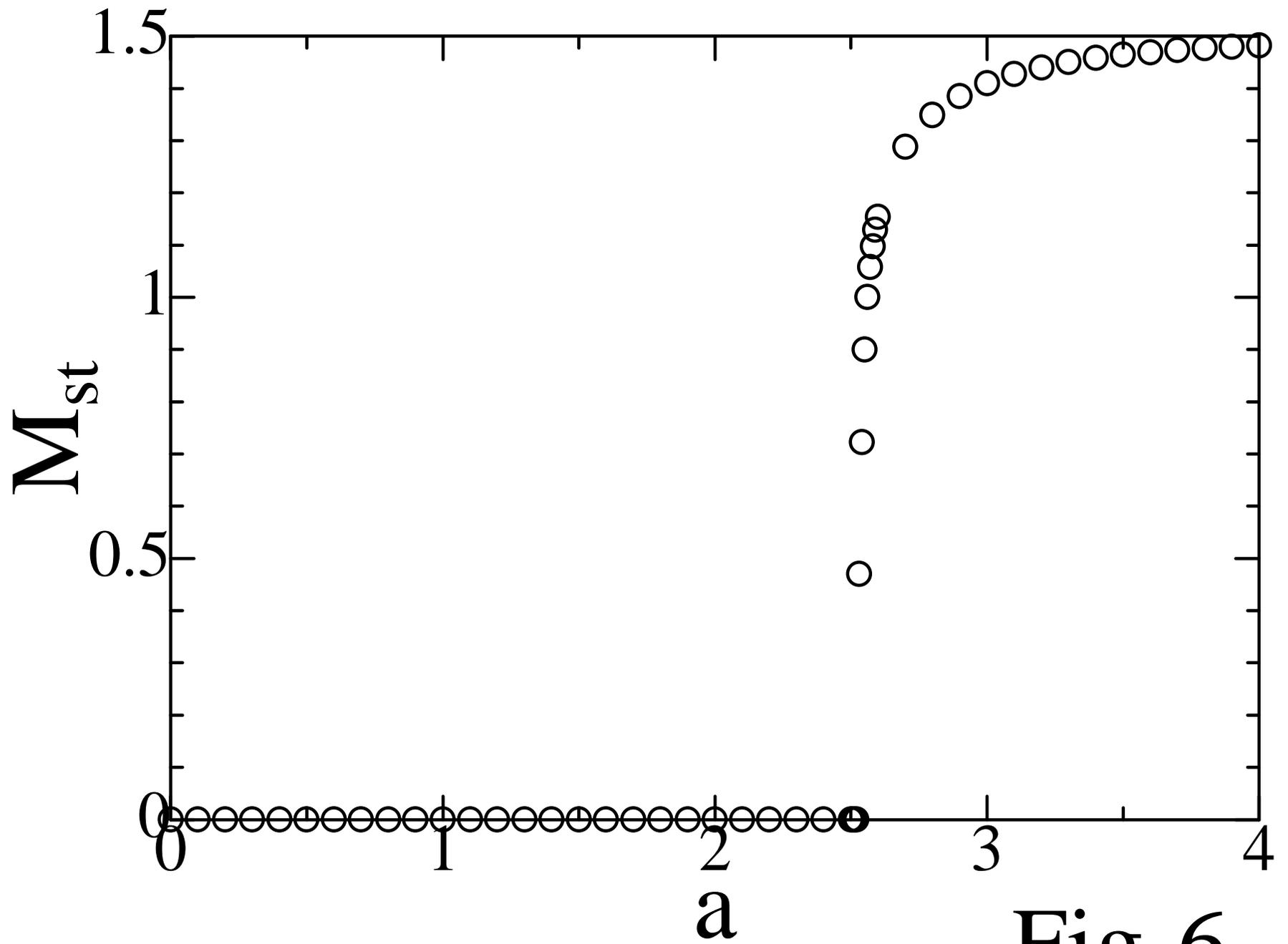
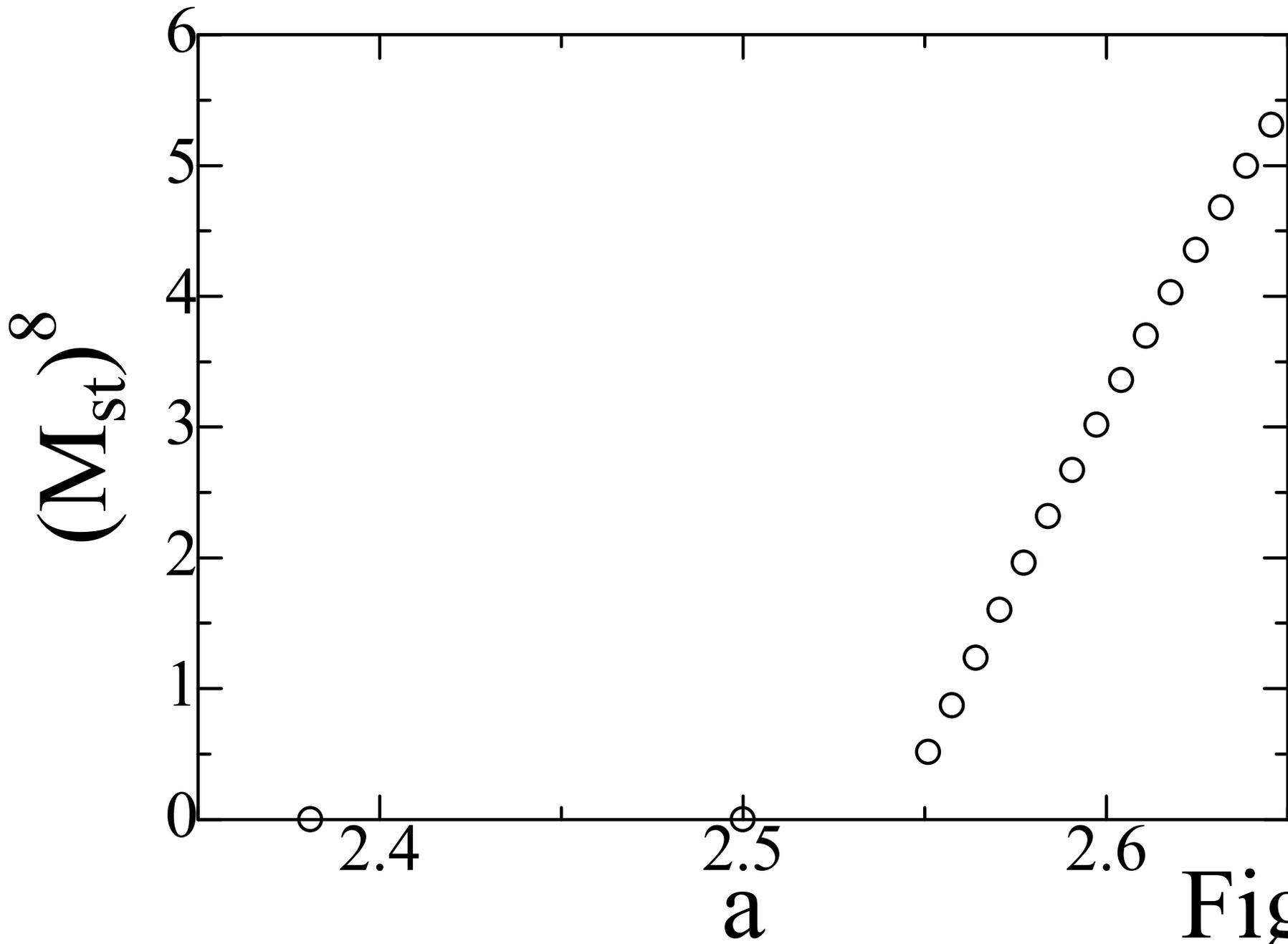
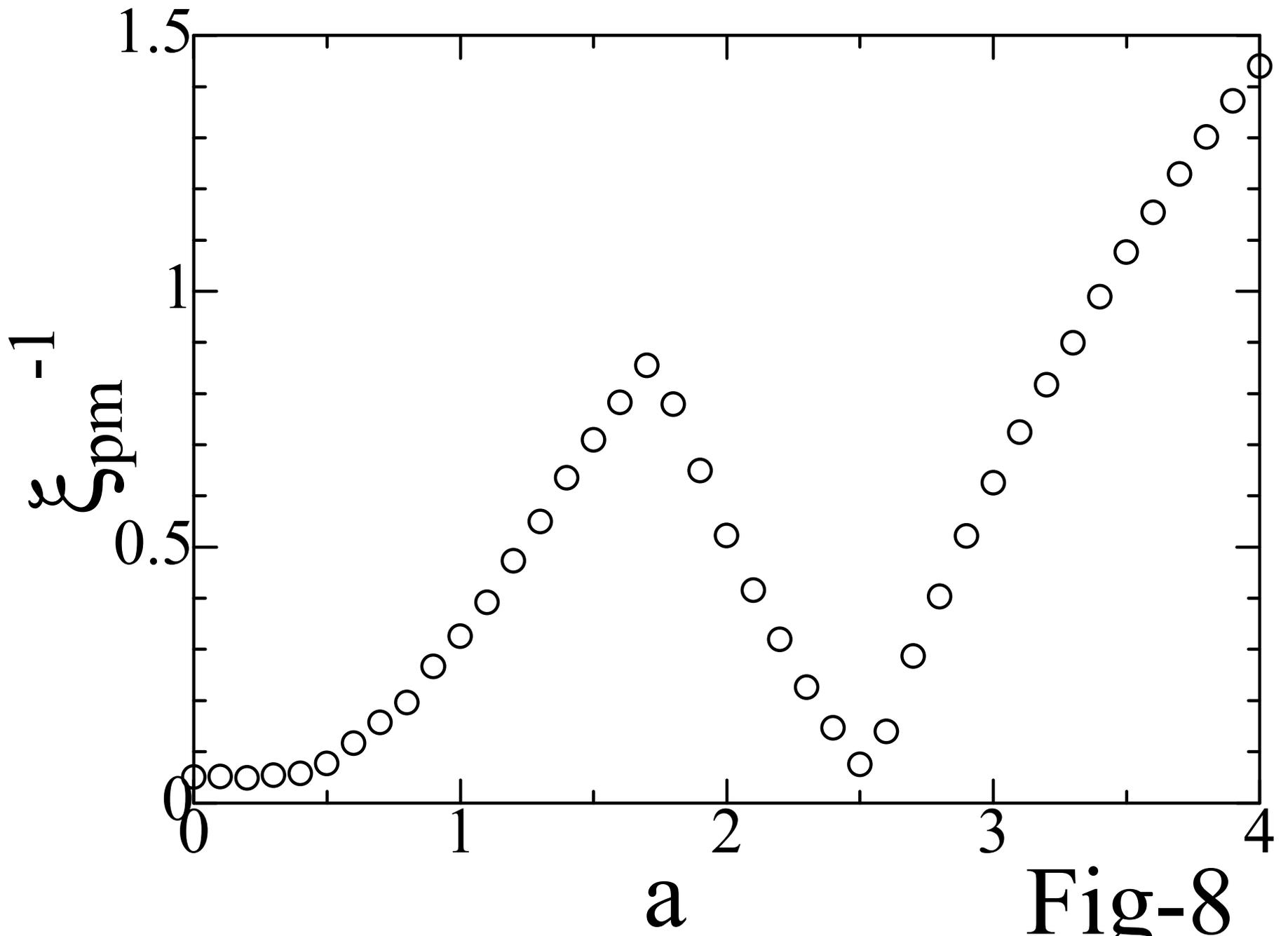
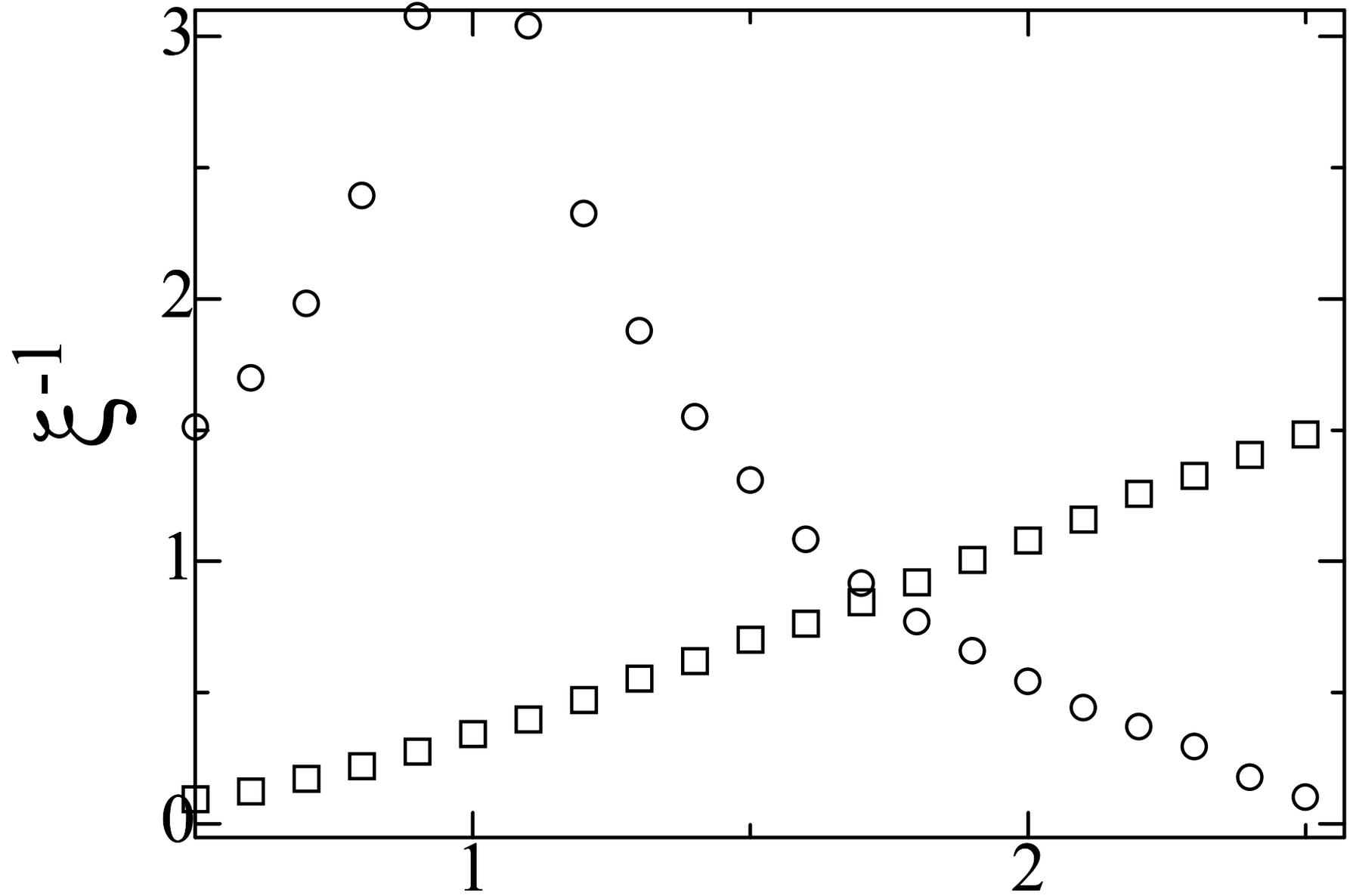
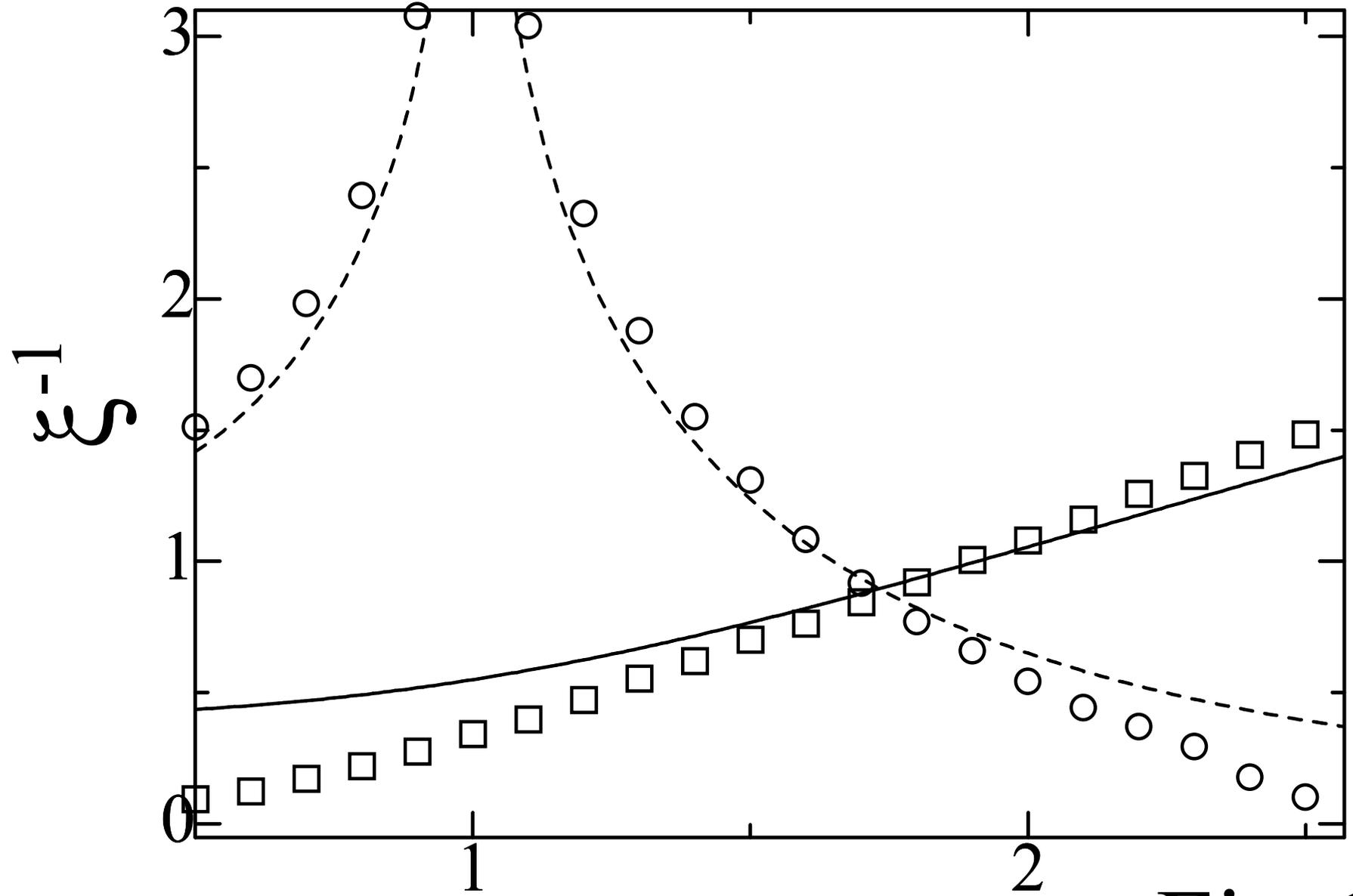
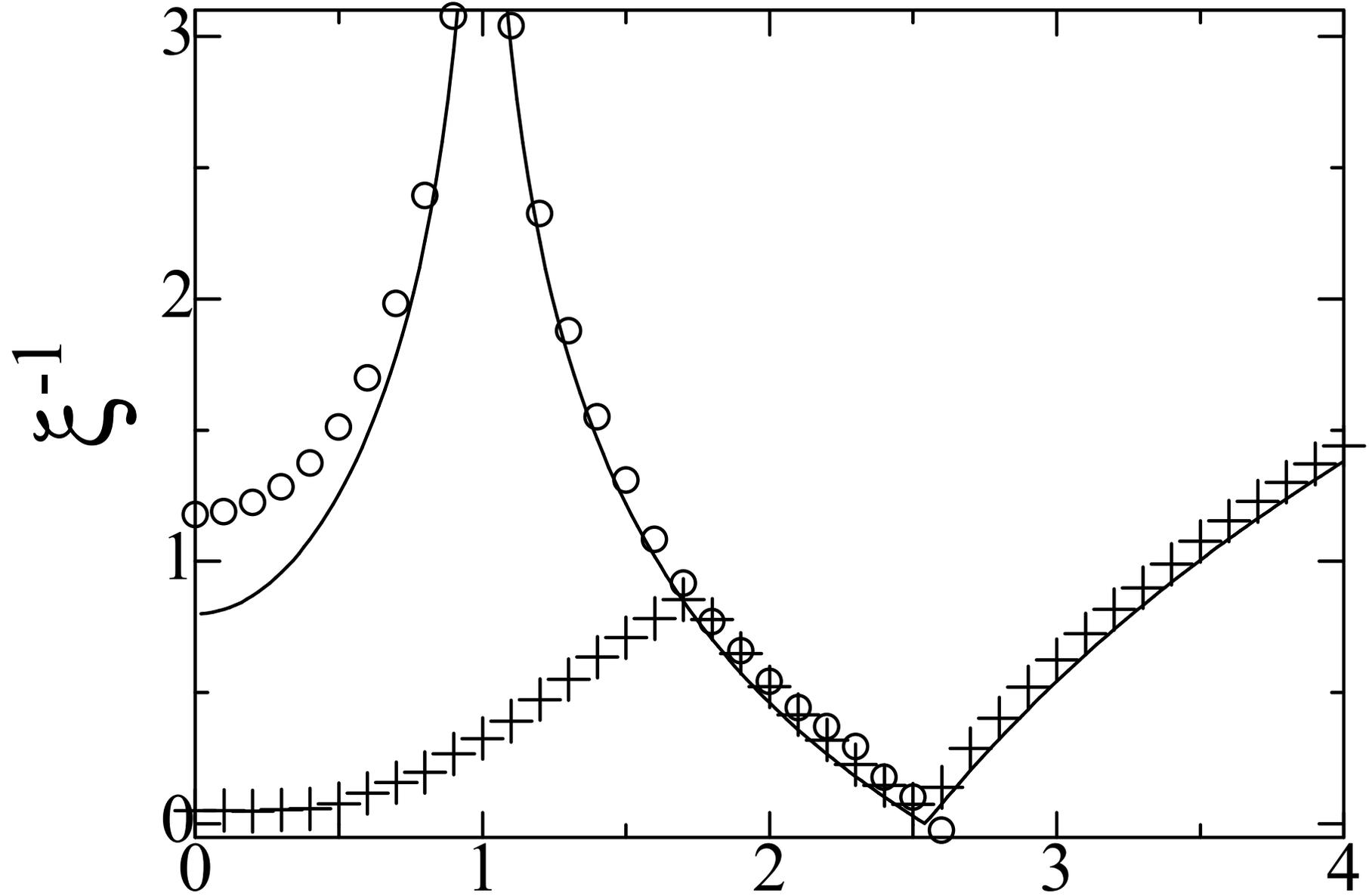

In Ref.~\cite{NKZ}, the ``asymptotic equivalence'' between the
deformed honeycomb VBS model in the large-$a$ region and the
free-fermion model is pointed out, which gives fairly accurate
estimation of the critical point $a_{c}$. As a further check of this
equivalence, we compare the correlation length $\xi_{\rm l}$ with that
of the free-fermion model which we denote by $\xi_{\rm ff}=\xi_{\rm
ff}(a)$.  The method given in
Refs.~\cite{Holzer-1,Holzer-2,Akutsu-Akutsu} allows us to obtain
$\xi_{\rm ff}(a)$ from the solution of $Q(i\omega^{*})=0$, where
$Q(\phi)$ is given by Eq.(24) in Ref.~\cite{NKZ} ($\xi_{\rm
ff}^{-1}=\omega^{*}$ in the disordered phase and $\xi_{\rm
ff}^{-1}=2\omega^{*}$ in the ordered phase). In Fig.\ref{fig:ff}, we
see a remarkable agreement in a unexpectedly wide parameter range
including the small-$a$ region where the ``asymptotic equivalence''
may not hold any longer.

\fbox{fig-11}

The cusp behavior of the correlation length implies that the isotropic
VBS model may be a ``singular'' point in the parameter space.  We
should point out that a similar cusp-like behavior of the correlation
length at the VBS point has been known in the $\beta-\xi^{-1}$
curve~\cite{Kennedy-2} and the $\theta-\xi$ curve~\cite{BBQ} of the
$S=1$ bilinear-biquadratic chain, where $-\beta$ and $\theta$ relate
to the relative amplitude of the biquadratic exchange term.  To
explore the relation between these cusps in the different parameter
spaces is an interesting problem, which may help us to characterize
the disordered phase in the 2D VBS-like models, in more detail: like
the string order parameter\cite{DenRom} which characterizes the
Haldane phase in the 1D case.

\section{Summary and Conclusion}
\label{sec5}
In this paper, we have presented a new way of application of the
 product wavefunction renormalization group (PWFRG) which is a variant
 of the density-matrix renormalization group (DMRG), to
 two-dimensional (2D) valence-bond-solid (VBS) type models.  The exact
 tensors-product-form (TPF) structure of the ground-state
 wavefunctions have allowed us to map the systems into two-dimensional
 classical statistical-mechanical models, namely, the vertex models
 which can be treated by the transfer-matrix method.  We extend the
 PWFRG method so that it can handle the asymmetric transfer matrix
 associated with the VBS-type models.

  For the isotropic 2D VBS models on the honeycomb lattice ($S=3/2$)
and the square lattice ($S=2$), we have confirmed that they are in the
disordered phase without the N\'{e}el order.  We have obtained
accurate values of correlation length for both models.  For the
anisotropically generalized (or deformed) VBS-type model on the
honeycomb lattice\cite{NKZ}, we have obtained detailed
parameter-dependence of the N\'{e}el order and the correlation length.
We have confirmed the anisotropy induced phase transition at a
critical value of the deformation parameter, and that this
second-order phase transition belongs to the 2D Ising-model
universality class.  In the disordered phase, we have found a
cusp-like behavior of the correlation length.  We have explained this
behavior as the crossover between the longitudinal and the transverse
correlation length.  Further, the PWFRG-calculated longitudinal
correlation length well agrees with the correlation length of the
free-fermion model in a fairly wide range of the deformation
parameter, which far exceeds the prediction of the ``asymptotic
equivalence''.\cite{NKZ}

 As has been mentioned in the Introduction, the present study is also
aimed at higher-dimensional generalization of the DMRG-type numerical
renormalization approach, based on the TPF-ansatz on the trial
wavefunctions.  From this point of view, what we have done in the
present paper is calculation of physical quantities under given
TPF-wavefunction with fixed local tensor.  What we should do next is
to vary the local tensor to make the variational calculation.  A
``direct numerical'' variational calculation can actually be done,
where we sweep the variational parameter and numerically find the
minimum of the energy-expectation. The PWFRG method is also useful for
reliable determination of the optimal variational value of the
parameter.  Studies along this line is now undertaken, whose details
will be published elsewhere.

\acknowledgments

One of the authors (Y. H.) would like to thank T. Nishino for
continuous encouragement.  This work was partially supported by the
Grant-in-Aid for Scientific Research from Ministry of Education,
Science, Sports and Culture (No.09640462), and by the ``Research for
the Future'' program of the Japan Society for the Promotion of Science
(JSPS-RFTF97P00201). One of the authors (Y. H.) is partly supported by
the Sasakawa Scientific Research Grant from The Japan Science Society,
and (K. O.) is supported by JSPS fellowship for young scientists.

\begin{figure}
\caption{
The pictorial representation of $A(\sigma|\{\eta\})$ of a $M$-valent
site.
}
\label{fig:A}
\end{figure}
\begin{figure}
\caption{
The mapping of the model on the honeycomb lattice to the model on the
 square lattice. Circles ($\bullet$), which express quantum spins
 ($S=3/2$), are connected by solid lines that form the honeycomb
 lattice.  The small letters ``a'' and ``b'' indicate sites of the two
 sublattices, respectively (see Sec.~\protect\ref{sec:dump-corr}).
}
\label{fig:map}
\end{figure}
\begin{figure}
\caption{
$E^p_q (a,b)$ is expressed according to Nishino's
 diagram~\protect\cite{Nishino}. Each circle ($\circ$) represents the
 mapped classical site, whose states are expressed by ``p'', ``q'',
 ``a'' and ``b''. These state variables take 4 states of the double
 index $(\eta^{i}_{\alpha},\xi^{i}_{\alpha})$.}
\label{fig:E}
\end{figure}
\begin{figure}
\caption{
The renormalized transfer matrix element
 $\widetilde{T}^{\alpha,i,j,\beta}_{\alpha',i',j',\beta'}$.
}
\label{fig:renT}
\end{figure}
\begin{figure}
\caption{
The direction along which correlation lengths of the model on the
honeycomb lattice are measured (a solid line).
}
\label{fig:CorrDirec-3}
\end{figure}
\begin{figure}
\caption{
The PWFRG result of the sublattice magnetization (staggered
magnetization) $M_{\rm st}$ as a function of the parameter $a$.
}
\label{fig:mz}
\end{figure}
\begin{figure}
\caption{
$(M_{\rm st})^8$ versus $a$. The linear behavior near
 $a_c (\protect\cong 2.54)$
 is seen.
}
\label{fig:mst8}
\end{figure}
\begin{figure}
\caption{
$\xi_{\rm pm}^{-1}$ as a function of $a$.
}
\label{fig:xipower}
\end{figure}
\begin{figure}
\caption{
The inverse of the longitudinal correlation length $\xi_{\rm l}^{-1}$
 ($\circ$) and the inverse of the transverse correlation length
 $\xi_{\rm t}^{-1}$ ($\Box$), in the disordered region.
}
\label{fig:xiinvzx}
\end{figure}
\begin{figure}
\caption{
Comparison of $\xi^{-1}$ for 1-dimensional ($S=3/2$) case and one for
2-dimensional case ($S=3/2$) in the disordered region.  Circles
($\circ$) and squares ($\Box$) are represented $\xi_{\rm l}^{-1}(a)$
and $\xi_{\rm t}^{-1}(a)$, respectively.  A solid curve is $(\xi_{\rm
t}^{(1D)}(a))^{-1}$ and broken curves are $(\xi_{\rm
l}^{(1D)}(a))^{-1}$.
}
\label{fig:xiinv}
\end{figure}
\end{document}